\documentclass[twocolumn,prl,preprintnumbers,showpacs,floatfix]{revtex4}

\usepackage{amsmath}
\usepackage{amsfonts}
\usepackage{amssymb}
\usepackage{graphicx}

\newcommand{\bra}[1]{\langle #1|}
\newcommand{\ket}[1]{|#1\rangle}
\newcommand{\braket}[2]{\langle #1|#2\rangle}
\newcommand{\I}{\mathrm{i}}
\newcommand{\E}{\mathrm{e}}
\newcommand{\rmi}{\mathrm{i}}
\newcommand{\rme}{\mathrm{e}}
\newcommand{\rmd}{\mathrm{d}}

\newcommand{\U}{\hat{\mathrm{U}}}
\newcommand{\Sc}{\hat{\mathrm{S}}}
\newcommand{\Parity}{\hat{\mathrm{P}}}
\newcommand{\Projector}{\hat{\Pi}}
\newcommand{\cm}{\text{c.m.}}
\newcommand{\rel}{\text{rel}}
\newcommand{\bg}{\text{bg}}

\newcommand{\res}{\text{res}}
\newcommand{\dte}{\text{DTE}}
\newcommand{\longit}{z}
\newcommand{\free}{\text{(0)}}
\newcommand{\Corr}{P}

\newcommand{\squezzit}{\enlargethispage*{\baselineskip}}

\begin{document}

\title {Bell test for the free motion of material particles} 

\author{Clemens Gneiting}
\affiliation{Arnold Sommerfeld Center for Theoretical Physics, Ludwig-Maximilians-Universit{\"a}t M\"unchen, Theresienstra{\ss}e 37, 80333 Munich, Germany}
\author{Klaus Hornberger}
\affiliation{Arnold Sommerfeld Center for Theoretical Physics, Ludwig-Maximilians-Universit{\"a}t M\"unchen, Theresienstra{\ss}e 37, 80333 Munich, Germany}
\preprint{published in: Phys.~Rev.~Lett.~{\bf 101}, 260503 (2008)}
\begin{abstract}
We present a scheme to establish non-classical correlations in the motion of two macroscopically separated massive particles without resorting to entanglement in their internal degrees of freedom. It is based on the dissociation of a diatomic molecule with two temporally separated Feshbach pulses generating a motional state of two counter-propagating atoms that is capable of violating a Bell inequality by means of correlated single particle interferometry. We evaluate the influence of dispersion on the Bell correlation, showing it to be important but manageable in a proposed experimental setup. The latter employs a molecular BEC of fermionic Lithium atoms, uses laser-guided atom interferometry, and seems to be within the reach of present-day technology.
\end{abstract}


\pacs{03.67.Bg, 37.25.+k, 03.65.Ud, 03.75.Gg}

\maketitle

\squezzit

{\it Introduction.---}The possibility of entangling macroscopically separate, non-interacting particles 
challenges our classical view of the world by putting into question the concepts of realism and locality 
\cite{Bell1987a}. 
Nowadays entangled states are routinely established with photons \cite{Brendel1999a}. 
Also the internal states of material particles have been entangled, e.g., using non-classical light as a carrier for the quantum correlations \cite{Julsgaard2001a,Sherson2006a,Matsukevich2006a,Moehring2007a,Rosenfeld2007akurz}, or 
their Coulomb interaction
in an ion trap \cite{Leibfried2005a,Haffner2005a}. 
However, the original discussion of entanglement focused on the \emph{motional} state of  massive particles, whose spatial separation is a dynamic feature of
the entangled two-particle wave function  \cite{Einstein1935a,Schrodinger1935a}.
The latter is spatially extended, unlike with internal entanglement, where the positions only play the passive role of separating the parties.
Since the positions and momenta are quantum observables with a direct classical analog, an observation of non-classical correlations  arising from macroscopically distinct phase space regions would therefore be a striking  demonstration of the failure of classical mechanics.

A convincing demonstration of non-classical correlations between two parties requires the experimental violation of a Bell-type inequality  \cite{Clauser1969a}. In the simplest case, it involves detecting dichotomic properties on each side, such as the polarization of spin-1/2 systems,
where a maximal violation is obtained if the spins are in a Bell state, say
$\ket{\Phi} = ({\ket{\uparrow}_1}  {\ket{\uparrow}_2} + \rme^{\rmi\phi}{\ket{\downarrow}_1} {\ket{\downarrow}_2})/\sqrt{2}$. 
Such a dichotomic property is not readily available in the motional state of free, structureless particles. One possibility is to consider observables which have no classical analogue, such as the displaced parity or pseudospin operators used to discuss entanglement in the `original EPR state' \cite{Banaszek1998a,Chen2002akurz}. While they are expedient for characterizing the quadratures of light fields, their experimental implementation seems exceedingly difficult in the case of free, macroscopically separated material particles, where only position measurements are 
easily realized. 

\begin{figure}
\includegraphics[width=0.95\columnwidth]{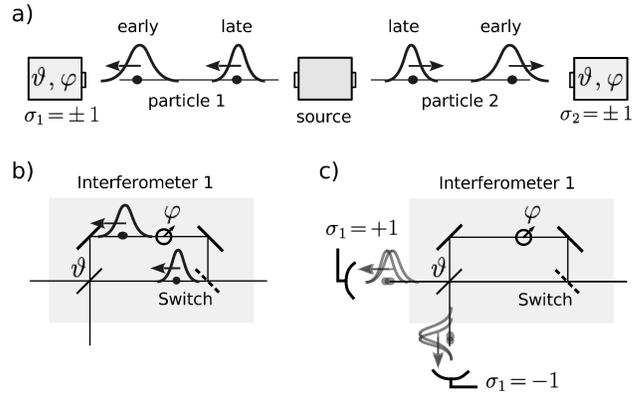}
\caption{\label{BellTest} 
(a) The two particles in a DTE  state  are each characterized by an early and a late wave packet component resulting from two different dissociation times. (b) Each particle is passed through an unbalanced Mach-Zehnder interferometer with a switch 
deflecting the early wave component into the long arm, while conducting the late component through the short arm. (c) The path difference is chosen such that the early and the late wave packets overlap. Detecting the particle in an output port, at a given phase $\varphi$ and splitting ratio $\vartheta$, amounts to a dichotomic measurement analogous to a spin $1/2$ detection in an arbitrary orientation. The existence of DTE is established, implying a macroscopically delocalized two-particle state, if the correlations at the single-particle interferometers violate a Bell inequality.
}
\end{figure}

Here we aim instead at producing a motional analogue of the Bell state $\ket{\Phi}$ and implementing a Bell test which requires a simple position measurement in the end.
The idea is to expose a diatomic molecule to a sequence of two temporally separated
dissociation pulses. Each of the two counter-propagating dissociated atoms then has an early and a late wave packet component corresponding to the two possible dissociation times, and their correlation may be called ``dissociation time entanglement'' (DTE), see Fig.~\ref{BellTest} (a).
If one regards
these components as spin state analogues in the motion (say, early
corresponds to spin up, late to spin down), the DTE state shows the same
structure as the Bell state $\ket{\Phi}$. Switched, unbalanced Mach-Zehnder interferometers on each  side then serve to mimic  arbitrary spin rotations, 
and the detection of each particle in one of the interferometer output ports completes the Bell test,  see Fig.~\ref{BellTest} (b,c).

The DTE state is a variant of the ``energy-time entangled'' state introduced in \cite{Franson1989a}, and it is closely related to the  ``time-bin entanglement'' of photons \cite{Brendel1999a,Tittel2000a,Simon2005a}, which has been used e.g. for establishing non-local correlations
over fiber distances of more than 50\,km \cite{Marcikic2004a} in a similar setup as in Fig.~\ref{BellTest}.
As a main difference, the DTE state 
is not composed of two single-particle product states, but it superposes their \emph{relative} coordinate, while the center-of-mass state  remains separable during the two-pulse dissociation process.

Our use of DTE reflects the necessity to come up with a state generation scheme appropriate to \emph{material} particles. Their finite mass and internal structure entail substantial complications which require a careful
investigation of whether a Bell violation can be expected at all.
Most prominently, one must account for the unavoidable wave packet dispersion, and only the recent progress in manipulating ultra-cold molecules (such as their condensation \cite{Kohler2006a} and controlled dissociation \cite{MukaiyamaDurrGreiner}) suggests the possibility to generate motional states that allow one to keep the detrimental effect of dispersion under control. We note that other ways to demonstrate nonlocal correlations of molecular dissociation products have been proposed in \cite{Opatrny2001a,Kheruntsyan2005a}.

We will show below that, by appropriately choosing a sequence of
magnetic field pulses, the Feshbach-induced dissociation of 
an ultracold ${}^6$Li$_2$ molecule within a guiding
laser beam can generate a motional state of macroscopically entangled atoms that is capable of violating a Bell inequality.
But first, to clarify the implications of dispersion, we determine the correlation function for a generic DTE state subject to 
correlated single-particle interferometry. 

\squezzit

{\it The general DTE Bell test.---}It is natural to take the bound molecular two-particle state to be separable in the center-of-mass ($\cm$)
and the relative ($\rel$) coordinate. 
Denoting by $\tau$ the period between the two dissociation pulses
and assuming the transverse motion to be frozen in the ground state of the guiding laser beam,
the longitudinal part of a pure DTE state then takes the form $\ket{\Psi_\dte} = (\U^\free_{z,\tau} \ket{\Psi_0} + \E^{\I \phi_{\tau}} \ket{\Psi_0})/\sqrt{2}$,
where 
$\ket{\Psi_0} = \ket{\psi_0^{\cm}}  (\ket{\psi_0^{\rel}}+\Parity\ket{\psi_0^{\rel}})/\sqrt{2}$.
It involves the free time-evolution operator $\U^\free_{z,t}$, the parity operator  $\Parity$, and a relative phase $\phi_{\tau}$ determined by the details of the two-pulse dissociation process.
The center-of-mass state of the original molecule  $\ket{\psi_0^{\cm}}$ is taken to rest in a wave guide, while $\ket{\psi_0^{\rel}}$ propagates into positive direction.

Indicating the  output ports of the two interferometers $j=1,2$ by $\sigma_j=\pm 1$, the immediate experimental outcome is characterized by the two-time probability density
$\mathrm{pr}(\sigma_1,\sigma_2;t_1,t_2)$ for detecting particles behind the respective ports at the times $t_1,t_2$. It depends on the phase settings $\varphi_j$ and mirror transmission angles $\vartheta_j$  of the interferometers, and the dispersive evolution of the DTE state leads to a complicated fringe pattern as a function of $t_1$, $t_2$, and $\tau$. However, as suggested by the above analogy with a discrete Bell test, a robust quantity characterizing entanglement is obtained by integrating the port-specific probabilities over all times, $\Corr_{\sigma_1,\sigma_2}=\int\mathrm{pr}(\sigma_1,\sigma_2;t_1,t_2)\rmd t_1\rmd t_2$. For any reasonable model of  the time-of-arrival detection, the correlation function can be equally evaluated by means of the projections $\Projector_{\sigma_1}\otimes \Projector_{\sigma_2}\equiv \Projector^{\sigma_1}_{\sigma_2}$ onto the (unbounded) spatial regions behind the respective ports, $\Corr_{\sigma_1,\sigma_2}= \lim_{t\to\infty} ||\Projector^{\sigma_1}_{\sigma_2}\U_{z,t}\ket{\Psi_\dte}||^2$.

Since we are only interested in the interferometer output, the effect of dispersion is best incorporated by time-dependent scattering theory, which separates the `raw action' of the interferometers, described by the S-matrices $\Sc_j$, from the free dispersive time evolution $\U_{z,t}^\free$. 
The projection of the DTE state component 
$\ket{\psi^\cm_0}\ket{\psi^\rel_0}$ 
onto a particular output-port combination $\sigma_1$, $\sigma_2$ then takes the form
\begin{equation} \label{eq:1}
\Projector^{\sigma_1}_{\sigma_2} \U_{z,t}^\text{(on)}
\ket{\psi^\cm_0}\ket{\psi^\rel_0} =  \U_{z,t}^\free 
[\Sc_{\sigma_1}^\text{(on)}\otimes \Sc_{\sigma_2}^\text{(on)}]
\ket{\psi^\cm_0}\ket{\psi^\rel_0} \,.
\end{equation}
Here we assume $t$ to be sufficiently large, so that the wave packets have passed the interferometers,  and we take the entrance switches to be in the ``on'' configuration, i.e., routing towards the long arms.
Implementing the phase shifts by the 
arm-length variations $\ell_j$, 
the projected S-matrices $\Sc_{\sigma_j}^\text{(on)}=\Projector_{\sigma_j}\Sc_j^\text{(on)}$ are given by
$\Sc^\textrm{(on)}_{\sigma_j = +1} = \exp(\I \hat{p}_j \ell_j/\hbar) \cos \vartheta_j$ and 
$\Sc^\textrm{(on)}_{\sigma_j = -1} = \exp(\I \hat{p}_j \ell_j/\hbar) \sin \vartheta_j$.
For the ``off'' configuration we have correspondingly
$\Sc^\textrm{(off)}_{\sigma_j = +1} = \sin \vartheta_j$ and
$\Sc^\textrm{(off)}_{\sigma_j = -1} = -\cos \vartheta_j$. 
Analogous relations hold for
$\ket{\psi_0^\cm}\Parity\ket{\psi_0^\rel}$.

The setup requires the dispersion-induced broadening of the wave packets  to remain much smaller than the separation between the early and the late components, so that the switches can be changed in between.
In this case the correlation function $\Corr_{\sigma_1,\sigma_2}$ can be evaluated by using Eq.~(\ref{eq:1}) and its variants even for non-pure and non-separable initial two-particle states,
$\ket{\Psi_0}\bra{\Psi_0}\to\varrho_0$.
One obtains
\begin{align} \label{eq:2}
\Corr_{\sigma_1,\sigma_2}(\ell_1,\ell_2) =&
\frac{1}{4} \Big\{ 1 + \sigma_1 \sigma_2 \; \text{Re} \Big[ \E^{- \I \phi_{\tau}}
\int \!\rmd p_1 \int \!\rmd p_2 
\nonumber\\ 
& \times 
\exp\Big(
\I \frac{\vec{p} \cdot \vec{\ell}}{\hbar}-\I \frac{\vec{p}^{\,2}\tau}{2 m \hbar} 
\Big)\mathrm{pr}(p_1,p_2) \Big] \Big\},
\end{align}
with $\vec{p} = (p_1,p_2)^{\text{T}}$, $\vec{\ell} = (\ell_1,\ell_2)^{\text{T}}$, and $\mathrm{pr}(p_1,p_2)=\bra{p_1,p_2}\varrho_0\ket{p_1,p_2}$ the momentum distribution function. 
For simplicity we take here the beam splitters to be symmetric 
($\vartheta_j= \pi/4$)  and the particles to have equal mass $m$.
Note that (\ref{eq:2}) is independent of the total time of flight and invariant under momentum phase transformations $\bra{p_1,p_2}\Psi_0\rangle\to \exp[\rmi\xi(p_1,p_2)]\bra{p_1,p_2}\Psi_0\rangle$, which includes spatial translations, thus rendering the correlation $\Corr_{\sigma_1,\sigma_2}(\ell_1,\ell_2)$ a robust signal.

For generic Gaussian states in the center-of-mass and relative motion Eq.~(\ref{eq:2}) can be evaluated in closed form.
The variances $\sigma^2_{p,\cm}$ and $\sigma^2_{p,\rel}$ then determine characteristic  dispersion times, 
$T_{\cm} = 2m \hbar/\sigma_{p,\cm}^2$  and $T_{\rel} = m \hbar/2\sigma_{p,\rel}^2$, indicating the transition to a dispersion-dominated spatial extension of the wave packets.
The expectation value of the relative momentum $p_{0,\rel}= m v_\rel/2 $ defines the reduced wave length $\lambdabar_{\rel} = \hbar/p_{0,\rel}$, which sets the scale for the non-local interference fringes in the explicit correlation function,
\begin{widetext}
\begin{align} \label{eq:3}
\Corr_{\sigma_1,\sigma_2}( \ell_1,\ell_2) = \frac{1}{4} \Bigg\{& 1 + \sigma_1 \sigma_2
\frac{\big( 1+{\tau^2}/{T_{\cm}^2} \big)^{-{1}/{4}}}{  \big( 1+{\tau^2}/{T_{\rel}^2} \big)^{{1}/{4}}}
\exp\bigg[
- \frac{T_{\rel}}{T_{\rel}^2+\tau^2}  \frac{(\ell_1-\ell_2-\tau v_{\rel})^2}{2 v_{\rel}\lambdabar_{\rel}}
- \frac{T_{\cm}}{T_{\cm}^2+\tau^2}\frac{(\ell_1+\ell_2)^2}{2 v_{\rel}\lambdabar_{\rel}}\bigg]
\nonumber \\
& \times \cos \bigg[ \frac{\ell_1-\ell_2}{\lambdabar_{\rel}}
+   \frac{\tau}{T_{\rel}^2+\tau^2} \frac{(\ell_1-\ell_2-\tau v_{\rel})^2}{2 v_{\rel}\lambdabar_{\rel}}
+  \frac{\tau}{T_{\cm}^2+\tau^2} \frac{(\ell_1+\ell_2)^2}{2 v_{\rel}\lambdabar_{\rel}} 
- \frac{\varphi_0}{2} \bigg] \Bigg\} ,
\end{align}
\end{widetext}\bigskip
with $\varphi_0 = \tau v_{\rel}/ \lambdabar_{\rel}+\arctan(\tau/T_{\cm})+\arctan(\tau/T_{\rel})+2\phi_{\tau}$.

\squezzit

These DTE correlations can violate a Bell inequality. This is seen from the structural similarity of (\ref{eq:3}) to the correlation function of the standard spin-1/2-based setup, 
$\Corr^{\text{spin}}_{\sigma_1,\sigma_2}(\varphi_1,\varphi_2) = \left\{ 1+\sigma_1 \sigma_2 \cos(\varphi_1-\varphi_2) \right\}/4$, where the  $\sigma_j=\pm 1$ denote the spin measurement  outcomes for analyzers tilted by the angles $\varphi_j$ with respect to the Bell state quantization axis. 
It follows from this analogy that an unambiguous demonstration of entanglement requires the fringe visibility of the correlation signal 
to exceed $1/\sqrt{2}$ over at least a few periods.

The dispersive suppression of this fringe visibility is described by those terms in (\ref{eq:3}) which depend on the characteristic times $T_{\cm}$ and $T_{\rel}$.
Specifically, the dispersion-induced distortion between
the early and the late wave packet components is reflected in the 
Lorentzian reduction factor and in the quadratic compression
of the fringe pattern, while their envelope mismatch causes the Gaussian suppression.
Based on this result one finds that nonlocal correlations can be observed, for $\lambdabar_{\rel}/(\tau v_{\rel}) \ll 1$, provided
$(1+\tau^2/T_\cm^2)(1+\tau^2/T_\rel^2)<4$.
In the following, we present a conceivable scenario for the generation of a DTE state, which meets these conditions.

{\it Experimental scenario for a DTE Bell test.---}We suggest to use a dilute molecular Bose-Einstein condensate (BEC)
produced from a 50:50 spin mixture of fermionic ${}^6\text{Li}$ as a starting point. 
It can be prepared efficiently and with near-perfect purity,
displaying  a huge lifetime of more than 10\,s
due to the Pauli blocking of detrimental 3-body collisions \cite{Jochim2003a}. A truly macroscopic time separation between the two pulses, say $\tau=1\,$s, is thus conceivable, and for a realistic dissociation velocity of $v_{\rel}=$1\,cm/s the de Broglie wave length  $\lambda_{\rel}=13.3\mu\text{m}$ is  large enough to pose viable stability requirements for the interferometers.

The BEC is prepared in a red-detuned, far off-resonant laser beam (transverse trap frequency $\omega_\text{G}/2\pi= 300$\,Hz), strong enough 
to guide the dissociated atoms towards the single particle interferometers. 
The intersection with a  second, weak laser beam  creates an elongated dipole trap for the BEC within the laser guide \cite{Fuchs2007a}. At the end of the preparation steps it is very shallow (depth $U_\text{T}/k_B=100\,\text{nK}$, $\omega_\text{T}/2\pi= 0.5$\,Hz) and only a small number of molecules (on the order of $10^{2}$) remains in the  BEC. These can be taken to be non-interacting, so that the initial longitudinal center of mass state $\ket{\psi_\text{T}}$ of the molecules is straightforwardly defined by the trap parameters.

Each interferometer (with path length difference $\tau v_\rel/2\simeq 5\,$mm) is implemented by two more red-detuned laser beams crossing the guide in a triangular arrangement at small angles. While the crossings act as beam splitters, the required atom mirror may be realized using an evanescent light field or a blue-detuned laser beam perpendicular to the interferometer plane \cite{Adams1994a,Kreutzmann2004a}. The time controlled appliance of such perpendicular blocking beams could also implement the switch. However, a simplified setup could replace the switch by an ordinary beam splitter, at the cost of 50\% post-selection. 

The fluorescence detection of the slow, strongly confined atoms at the two output guides can be done with near unit efficiency and with single particle resolution, since no particular spatial or temporal accuracy is needed. 
This single-particle resolution is crucial since 
events where more than one molecule gets dissociated are disregarded in the present scenario, relying on post-selection. In a more refined setup, it is conceivable to use a specially prepared optical lattice where each site is occupied by at most one molecule \cite{Volz2006a}.

All this implies that the molecular dissociation in presence of the wave guide must meet a number of rather restrictive criteria to render a demonstration of macroscopic entanglement possible.
Only the transverse ground state of the guide may be populated 
to admit the above quasi-one-dimensional description of the interferometers, while both the momentum spread of the wave packets and the dissociation probability must be sufficiently small. In order to judge the feasibility of our experimental scenario, we now discuss how the dissociated part of the state, $\ket{\Phi_{\bg}(t)}$,  depends on the magnetic field pulse sequence and the resonance parameters.

A  Green function analysis within the two-channel single-resonance approach shows that after an arbitrary magnetic field pulse sequence 
(close to an isolated resonance)  $\ket{\Phi_{\bg}(t)}$ is described, for low energies, at positions far from the dissociation center, and at large times, by the asymptotic form $\ket{\Phi_{\bg}(t)} \sim C_{\bg}  \ket{\varphi_{0,0}^\cm}\ket{\varphi_{0,0}^\rel}\U^\free_{\longit,t}\ket{\Psi_{\longit}}$, where $\U^\free_{\longit,t}$ is the free propagator in the longitudinal direction and where the transverse motion is frozen in the  harmonic ground state, $\ket{\varphi_{0,0}^\cm}$ and $\ket{\varphi_{0,0}^\rel}$, respectively, of the guiding laser beam. The longitudinal state is determined by
$\braket{p_\cm,p_\rel}{\Psi_{\longit}} = \tilde{C}(p_\cm^2/4m\hbar+p_\rel^2/m\hbar+2\omega_\text{G})\bra{p_\cm}\psi_\text{T}\rangle/||\tilde{C}||$, where $\tilde{C}(\omega)$ is the Fourier transform 
of the  closed channel probability amplitude $C(t)$
and $||\tilde{C}||^2 = \int \rmd p_\cm \rmd p_\rel |\tilde{C}(p_\cm^2/4m\hbar+p_\rel^2/m\hbar+2\omega_\text{G})|^2 |\bra{p_\cm}\psi_\text{T}\rangle|^2$.
${C}(t)$, in turn, is determined by the coupled channel dynamics as induced by the externally controlled magnetic field $B(t)$.
The dissociation probability is  given by
$|C_{\bg}|^2 = \omega_{\text{G}} a_{\bg} \mu_{\res} \Delta B_{\res} ||\tilde{C}||^2 / \pi \hbar^2$.
It involves the background scattering length  $a_{\bg}$, the resonance width  $\Delta B_{\res}$, and $\mu_{\res}$, the difference between the magnetic moments of
the resonance
state and the open channel.

\squezzit

While the association of the molecules is best done at a broad resonance \cite{Jochim2003a}, shifting to a narrow resonance (e.g., $\Delta B_{\res}=1$\,mG, $\mu_{\res}=0.01\,\mu_\text{B}$, $a_{\bg}=100 a_0$) one can ensure by choosing field pulses $B(t)$ with short duration (e.g. $T=60\,$ms) that about a single molecule dissociates on average. 
A sequence of two square pulses with base value $B_0$ and height $\Delta B$  sweeping over the resonance position $B_{\res}$ 
then generates a DTE longitudinal wave packet of the required form,
$\ket{\Psi_{\longit}} =  [\U^\free_{\longit,\tau}  + \E^{\I \phi_\tau}]  \ket{\Psi_0}/\sqrt{2}$.
The momentum distribution is given by
$|\bra{p_\cm,p_\rel}\Psi_0\rangle|^2=p_0\bar{p}^4\text{sinc}^2[(p_\cm^2/4+p_\rel^2-p_0^2)/\Delta p^2]|\bra{p_\cm}\psi_\text{T}\rangle|^2/\pi[(p_\cm^2/4+p_\rel^2-p_0^2+\bar{p}^2)\Delta p]^2$, where we define $p_0^2/m=\mu_\text{res}(B_0+\Delta B-B_\text{res})-2U_\text{T}-\hbar\omega_\text{G}$, as well as $\bar{p}^2/m=\mu_\res\Delta B$ and $\Delta p^2=2m\hbar/T$ \footnote{The normalization of the momentum distribution assumes $\Delta p,\sigma_{p,\cm} \ll p_0$, which holds for our parameters.}. 
Instead of evaluating Eq.~(\ref{eq:2}) directly with this momentum distribution, it is more transparent to approximate the latter by Gaussians, which allows one to apply the analysis following Eq.~(\ref{eq:3}).
For reasonable pulses $B_0+\Delta B-B_\text{res}=350$\,mG these Gaussians are centered at $p_{0,\rel}=\pm m v_\rel/2$ and $p_{0,\cm}=0$, with spreads of $\sigma_{p,\rel}=1.196 m \hbar/(p_0 T)$ and $\sigma_{p,\cm}=\sqrt{\hbar\omega_\text{T}m/2}$, respectively. This yields $T_{\rel} = 3.4$\,s and $T_\cm=0.64\,$s, implying a visibility of about 72\% in the correlation signal, which exceeds the threshold value of $1/\sqrt{2}$.

Our analysis thus shows that the observation of a macroscopic DTE state is feasible with material particles,  even though dispersion poses tight
constraints. The technological challenge is substantial, but not insurmountable. Stable lasers are required and the magnetic pulse sequence must be reproducible with a relative accuracy of  $10^{-5}$ from shot to shot, so that the relative phase between the early and late components, given by 
$\phi_\tau \simeq [2 U_\text{T}\tau-\mu_{\res} \Delta B \, T + \mu_{\res} (B_{\res}-B_0) \tau]/\hbar+\omega_\text{G}\tau$, varies less than 50\,mrad. 
Realistic choices of the laser wave length ($1\,\mu$m) and the vacuum pressure ($10^{-8}\,$mbar) suffice to suppress decoherence due to scattering of off-resonant photons or background gas particles.

A great advantage  of this setup is that no interferometric stability is required between the two interferometers, so that truly macroscopic separations are feasible. 
Moreover, the DTE state 
reveals its entanglement robustly since neither a spatial nor temporal resolution is required in the detection.

This work was supported by the DFG Emmy Noether program.


\begin{thebibliography}{10}

\bibitem{Bell1987a}
J.~S. Bell,
\newblock {\em Speakable and Unspeakable in Quantum Mechanics} (Cambridge
  University Press, 1987).

\bibitem{Brendel1999a}
J.~Brendel, N.~Gisin, W.~Tittel, and H.~Zbinden,
\newblock Phys. Rev. Lett. {\bf 82}, 2594 (1999).

\bibitem{Julsgaard2001a}
B.~Julsgaard, A.~Kozhekin, and E.~S. Polzik,
\newblock Nature {\bf 413}, 400 (2001).

\bibitem{Sherson2006a}
J.~Sherson {\em et~al.},
\newblock Nature {\bf 443}, 557 (2006).

\bibitem{Matsukevich2006a}
D.~N. Matsukevich {\em et~al.},
\newblock Phys. Rev. Lett. {\bf 96}, 030405 (2006).

\bibitem{Moehring2007a}
D.~Moehring {\em et~al.},
\newblock Nature {\bf 449}, 68 (2007).

\bibitem{Rosenfeld2007akurz}
W.~Rosenfeld {\em et~al.}, \newblock Phys. Rev. Lett. {\bf 98}, 050504 (2007).

\bibitem{Leibfried2005a}
D.~Leibfried {\em et~al.},
\newblock Nature {\bf 438}, 639 (2005).

\bibitem{Haffner2005a}
H.~Haffner {\em et~al.},
\newblock Nature {\bf 438}, 643 (2005).

\bibitem{Einstein1935a}
A.~Einstein, B.~Podolsky, and N.~Rosen,
\newblock Phys. Rev. {\bf 47}, 777 (1935).

\bibitem{Schrodinger1935a}
E.~Schr{\"o}dinger,
\newblock Proc. Camb. Phil. Soc. {\bf 31}, 555 (1935).

\bibitem{Clauser1969a}
J.~F. Clauser, M.~A. Horne, A.~Shimony, and R.~A. Holt,
\newblock Phys. Rev. Lett. {\bf 23}, 880 (1969).

\bibitem{Banaszek1998a}
K.~Banaszek and K.~Wodkiewicz,
\newblock Phys. Rev.~A {\bf 58}, 4345 (1998).

\bibitem{Chen2002akurz}
Z.-B. Chen {\em et~al.}, \newblock Phys. Rev. Lett. {\bf 88}, 040406 (2002).

\bibitem{Franson1989a}
J.~D. Franson,
\newblock Phys. Rev. Lett. {\bf 62}, 2205 (1989).

\bibitem{Tittel2000a}
W.~Tittel, J.~Brendel, H.~Zbinden, and N.~Gisin,
\newblock Phys. Rev. Lett. {\bf 84}, 4737 (2000).

\bibitem{Simon2005a}
C.~Simon and J.~P. Poizat,
\newblock Phys. Rev. Lett. {\bf 94}, 030502 (2005).

\bibitem{Marcikic2004a}
I.~Marcikic {\em et~al.},
\newblock Phys. Rev. Lett. {\bf 93}, 180502 (2004).

\bibitem{Kohler2006a}
T.~K{\"o}hler, K.~G{\'o}ral, and P.~Julienne,
\newblock Rev. Mod. Phys. {\bf 78}, 1311 (2006).

\bibitem{MukaiyamaDurrGreiner}
T.~Mukaiyama {\em et~al.}, \newblock Phys. Rev. Lett. {\bf 92}, 180402 (2004); 
S.~D{\"u}rr, T.~Volz, and G.~Rempe,
\newblock Phys. Rev.~A {\bf 70}, 031601 (2004);
Greiner {\em et~al.}, \newblock Phys. Rev. Lett. {\bf 94}, 110401 (2005).

\bibitem{Kheruntsyan2005a}
K.~V. Kheruntsyan, M.~K. Olsen, and P.~D. Drummond,
\newblock Phys. Rev. Lett. {\bf 95}, 150405 (2005).

\bibitem{Opatrny2001a}
T.~Opatrn{\'y} and G.~Kurizki,
\newblock Phys. Rev. Lett. {\bf 86}, 3180 (2001).

\bibitem{Jochim2003a}
S.~Jochim {\em et~al.},
\newblock Science {\bf 302}, 2101 (2003).

\bibitem{Fuchs2007a}
J.~Fuchs {\em et~al.},
\newblock J.~Phys.~B: At. Mol. Opt. Phys. {\bf 40}, 4109 (2007).

\bibitem{Adams1994a}
C.~S. Adams, M.~Siegel, and J.~Mlynek,
\newblock Phys. Rep. {\bf 240}, 143 (1994).

\bibitem{Kreutzmann2004a}
H.~Kreutzmann {\em et~al.},
\newblock Phys. Rev. Lett. {\bf 92}, 163201 (2004).

\bibitem{Volz2006a}
T.~Volz {\em et~al.},
\newblock Nature Physics {\bf 2}, 692 (2006).

\end{thebibliography}

\end{document}